\documentstyle[12pt,aasms4]{article}

\begin{document}

\def\etal{{\it et al.\/\ }}
\def\today{\ifcase\month
           \or January   \or February \or March    \or April 
           \or May       \or June     \or July     \or August
           \or September \or October  \or November \or December\fi
           \space\number\day, \number\year}
\def\ga{\lower 2pt \hbox{$\, \buildrel {\scriptstyle >}\over{\scriptstyle \sim}\,$}}
\def\la{\lower 2pt \hbox{$\, \buildrel {\scriptstyle <}\over{\scriptstyle \sim}\,$}}

\title{\bf RESULTS FROM A SURVEY OF GRAVITATIONAL MICROLENSING TOWARDS M31}

\author{Arlin P. S. Crotts$^{1,2}$ and Austin B. Tomaney$^{1,2}$}

\affil{Department of Astronomy, Columbia University, 538 W.~120th St., 
New York, NY~~10027}

\affil{\bigskip
$^1$ Based on observations at the Vatican Advanced Technology Telescope
(The Alice P. Lennon Telescope and Thomas J.  Bannan Astrophysical Facility),
Mt Graham Arizona}

\affil{\bigskip
$^2$ Based on observations at the Kitt Peak National Observatory (KPNO),
National Optical Astronomy Observatories, operated by AURA, Inc.~for the NSF}

\authoremail{arlin@eureka.phys.columbia.edu, austin@odyssey.phys.columbia.edu}

\begin{abstract}

We describe the results of a search for microlensing events affecting stars in
the outer bulge and inner disk of M31, due both to masses in M31 and the
Galaxy.
These observations, from 1994 and 1995 on the Vatican Advanced Technology
Telescope and KPNO 4m, are sufficient to rule out masses in the range of
$\sim 0.003~M_\odot$ to 0.08~$M_\odot$ as the primary consistuents of the
mass of M31 towards this field.
Furthermore we find six candidate events consistent with microlensing due to
masses of about $1~M_\odot$, but we suspect that some of these may
be cases where long-period red supergiant variables may be mistaken for
microlensing events.
Coverage from anticipated data should be helpful in determining if these
sources maintain a constant baseline, and therefore are best described by
microlensing events.

\end{abstract}


\clearpage

\section{INTRODUCTION}

One of the most significant and stubborn mysteries in astrophysics today
concerns the nature of the dark matter in spiral galaxies (e.g.~Rubin et
al.~1978).
The least radical candidate for such dark matter is baryonic objects which
are too large to be detected as dust or gas.
Indeed, if the Hubble constant is not too large, a significant fraction of the
baryons, as implied by Big Bang nucleosynthesis, must be hidden as dark matter
(Walker et al.~1991).
Arguments have been made for why such dark baryonic objects are unlikely on
individual mass scales of atoms to brown dwarfs (Hills 1986, Hegyi \& Olive
1986).
Still, objects of primordial composition and more massive than about $10^{-7}
M_{\odot}$ might be expected to resist evaporation until the present day
(de R\'ujula et al.~1992), while masses smaller than about $0.077 ~ M_\odot$
would fail to ignite as stars (Burrows et al.~1993).

Astrophysicists' frustration explaining the dark matter with any directly
detectable objects has led to the suggestion that gravitational microlensing
might be used to at least betray the presence of individual objects via their
effects on background stars as sources (Paczynski 1986), and thereby give some
indication as to their mass.
Such searches have recently taken place towards the Large Magellanic Cloud
(LMC) (Alcock et al.~1996, Aubourg et al.~1995) and Bulge (Alcock et al.~1995,
Paczynski et al.~1994), with searches towards the LMC ruling out most of the dark
matter being composed of substellar-mass objects (Aubourg et al.~1995, Alcock
et al.~1996) heavier than about $10^{-6} M_\odot$, while suggesting that a
large fraction might have the same component mass as low-mass stars (Alcock et
al.~1996).
Given the uncertainty of the Galactic halo's distribution of MAssive, Compact,
Halo Objects (MACHOs) and therefore the lensing geometry leading to events,
the relationship between mass and observed microlensing lightcurve timescale is
still unclear.

In part because of its unique geometry with respect to Earth and partially due
to high predicted optical depths ($\tau$) due to lensing, M31 is a uniquely
powerful venue for studying microlensing.
Early we realized that an M31 microlensing survey would show particular
advantages if the practical aspects of studying such a distant, crowded field
of stars could be overcome.

We found such an approach, briefly outlined by Crotts (1992) with a complete
description of the realistic technique and preliminary results found in Tomaney
and Crotts (1996, hereafter TC).
By subtracting images in a time sequence, then performing ``difference image
photometry'' (DIP, also know as ``pixel lensing''), we can study the residual
point sources due to variables, while the signals from the many crowded,
non-varying stars subtract away.
With a practical method of observation and analysis, we can exploit the
advantages inherent in studying M31: 1) very small component mass limits, due
to the small angle subtended by the photosphere of M31 stars compared to the
Einstein radius of objects of solar mass (c.f. TC for low-mass results), 2) the
ability to study different parts of M31, thereby studying the spatial
distribution of microlensing objects, 3) the ability to study many stars at
once in fields of high $\tau$, thereby detecting events in short periods of
observation, and 4) the constrained microlensing geometry, due to the fact that
lensing mass is concentrated over the center of the galaxy, thereby allowing a
better determination of the MACHO mass given microlensing event timescale.

It is our hope that by studying M31 in this way, both its halo and bulge,
one can more readily understand both these results and those obtained in the
Galaxy.
This paper presents our results from our first season of observation toward
this goal.

\section{Observations}

Observations were made primarily at the Vatican Advanced Technology
Telescope on 31 December 1994, 31 January, 17 October through 4 November and
20 November through December 3, 1995, using an imager especially made for
accommodating DIP.
Additional observations were made with the prime focus KPNO 4-m Prime Focus
Camera on 24-27 September 1994 and 28 August 1995.

The VATT data were taken using field centers of 00:43:16.5 +41:11:33 and
00:42:13.4 +41:20:44 (J2000), and rotation of the CCD so that its sides were
roughly parallel to M31's principle axes.
The first field looks past the bulge, intercepting the far side of the disk
along the minor axis.
Most data were obtained in the first field, such that light curves in the
second are too incomplete to treat here.
Observations on the KPNO 4m were centered close to the first VATT location,
such that the 4m field (16.4 arcmin on a side) encompasses all but 0.1\% of the
VATT field (11.3 arcmin on a side).
The two telescopes' fields are rotated 38$^\circ$ with respect to each other.
We do not consider here 4m data falling outside the VATT field.
Assuming a distance to M31 of 770~kpc (4.46 arcmin/kpc), the VATT field covers
a range of 0.4 to 2.9~kpc along the minor axis, which projects to 1.8 to
13.1~kpc along the disk, assuming an inclination $i=77^\circ$.
We use the filter bands described in TC, essentially broad R and I bands to
match the fact that most of our target stars in M31 are red giants or
supergiants and therefore brighter in redder bands.

\section{Analysis}

The process of difference image photometry 
consists of careful flat-fielding, coordinate registration and
photometric scaling of the data, followed by point spread function (PSF)
matching between frames (detailed in TC).
Using a convolution kernel approximating the quotient between PSFs in the image
from a particular epoch and that from a high-$S/N$, good-seeing stack of many
images, the PSFs are matched either by degrading the image stack or single
epoch's image to match the other.
In the case of the VATT data, the entire frame can be corrected in this way due
to the optical design of our imager, which insures that a single convolution
kernel is need over the entire (or nearly all) of the image (see TC for more
details).
This is especially easy because in these data we deal only with nightly sums,
composed of $\approx 5-20$ similar exposures, so that individual irregularities
of single exposures, such as guiding errors, average out.
After PSF matching is accomplished, the stack (taken from many night's data)
and the individual night's image are subtracted, leaving a field of noise at
nearly the photon shot noise level, as well as isolated positive or negative
point source residuals due to variable stars.
An example of this is shown in Figure 1, for the fourth candidate microlensing
event detailed in Figure 2.
DIP is completed by performing aperture photometry on these isolated residual
sources, which can then be incorporated into light curves.

The error bars presented with the lightcurves in Figures 2 and 3 show the
fluctuations in the difference image on the scale of the PSF in regions
adjacent to each residual source.
Sources were catalogued by requiring at least a 4$\sigma$ detection in at least
two nightly sums (or 6$\sigma$ in the 24\% of the image containing the bulge
and closest to the minor axis), then tracing the lightcurve by aperture
photometry in other epochs at the same location.
In future papers we will track additional sources by 1) sampling variations on
sub-night scales, and 2) summing difference images so that weaker residuals can
be tracked over longer timescales.
Even without these refinements, however, we locate over 2000 sources within the
VATT field.

\section{Results}

The results of this construction of lightcurves from nightly sums is that no
source is only on two consecutive nights, and that none of the sources seen,
with one exception, is consistent with microlensing events on any but nearly
the longest timescales sampled by our survey.
In the latter cases, we portray the lightcurves of the six candidate events in
Figures 2, and other information in Table 1, including their positions (J2000)
and distance along M31's minor axis ($d$).
Assuming that they are microlensing events, several other parameters can also
be extracted: the duration (Einstein radius crossing time $t_e$), lensing
impact parameter (normalized to the Einstein radius: $u_o = u/R_e$), and
source baseline magnitude ($R$).
Not given are the two other fit parameters, time of peak amplification and
flux zero-point offset due to image subtraction.
Additionally we give the goodness of the best lensing fit (for point sources
and masses), and the most probable mass of the lens.
It appears that fit residuals are slightly larger than expected from
photometric measurement error alone, seen particularly as a surplus in the
number of 3$\sigma$ or greater residuals, which are inconsistent with
neighboring points.
One possibility for this noise is underlying RR Lyrae variables, which should
be evident at the 1-2$\sigma$ level, either coincident with the source or in
its photometric background annulus.
We will investigate this problem further in Tomaney et al.~(1996).
We stress that {\it we do not claim that these are microlensing events at least
until their lightcurves are observed to fall and remain at the pre-event
baseline} during the 1996 observing season or thereafter.

One reason for our caution is the lightcurve shown in Figure 3 for the variable
star found at 0$^h$43$^m$37.$^s$9 $+$41$^\circ$14$^\prime$57$^{\prime\prime}$
(J2000), 1.92~kpc from the major axis.
It is fit well by a microlensing lightcurve during its rise and fall, but does
not maintain a consistent baseline before and after the event.
Upon inspection of lightcurves of Mira-type variables (Wesselink 1987), we find
a small fraction whose lightcurves around maximum light mimic
the behaviour of microlensing lightcurves.
A sparse sampling of points beyond maximum light, if chosen unfortuitously,
might fail to distinguish such a variable from a microlensing event.
We find several such variables in our VATT field.
Further reasons for suspicion is the similarity in timescales to those of miras
(except the first event), and similar shapes, indicated by $u_o$ values
which cluster around 0.6 (except for the second event).
Additionally, it is strange that all sources have $R\approx21$, close to the
magnitude that would correspond to a mira pulsation (given the inferred $u_o$),
but brighter than what we might expect for lensed sources given the luminosity
function of stars in the field.
We suspect that several of these events are not due to microlensing at all, but
might be associated with bright variable stars.
Another season of observation, which we plan, will determine if these sources
maintain a constant baseline and are therefore likely to be lensed.

The reality of these events can be tested in terms of the distribution of $u_o$
values via a one-sample Kolmogorov-Smirnov (K-S) test.
The theoretical distribution is derived using a luminosity function in $R$
$\phi \propto 10^{\alpha R}$, where $\alpha = 0.59$ best describes the behavior
of star counts and surface brightness fluctuations in the field, and agrees
with other works (Tomaney \& Crotts 1996, and references therein).
Very low $u_o$ values ($u_o \la 0.02$) are not realistic since high
amplification events have timescales too short for us to detect; high $u_o$
events ($u_o \ga3.5$) have amplifications too subtle for us to detect, as well.
We take as the $u_o$ upper bound the lesser of the above upper bound and the
maximum value providing sufficient amplification to reach our flux threshold
for a given magnitude.
For values of $u_o$ between these limits, we assume a uniform distribution of
events in $u_o$, at a given magnitude.
The largest value of the K-S distance $D$ occurs at the smallest observed $u_o=
0.369$, due to the lack of small $u_o$ events, and has a value $D\approx0.7$.
Assuming that all six candidates are true microlensing events, the null
hypothesis (consistency with microlensing) is rejected at the 99.5\% level.
If half the candidates are microlensing events (and the minimum still
$u_o=0.369$), the null hypothesis is rejected at approximately the 90\% level.
It is unlikely that all of the events are due to microlensing, but this test
cannot rule out that a large fraction may be.

These caveats aside, {\it if} these events are microlensing events, then we can
say several things about them.
The first and second events land in the bulge-dominated region, and hence 
likely involve bulge sources.
The third and fourth might be due to disk sources (but have a high probability
of belonging to the bulge) and also rest in the region where bulge lenses
may dominate over halo lenses.
The fifth and sixth events, if genuine, might easily be halo lenses acting on
disk sources.
In the case of third through sixth events, the most probable source-lens
distance is $d/cos~i$, allowing us to compute a most likely mass, given $t_e$.
(We assume a disk rotation speed of 260~km~s$^{-1}$ [Braun 1991] and a
halo/bulge velocity dispersion of 160~km~s$^{-1}$ [Kent 1989], of which
$\sqrt{2/3}$ is in the transverse direction.
Earth's transverse motion is negligible.) 
We assume no rotation of the bulge; it could be as large as
$\sim$100~km~s$^{-1}$ in our field (Kent 1989), meaning that inferred masses
might tend to split into a bimodal distribution of under- and overestimated
values, with peaks differing by as much as a factor of two in timescale, or
four in mass.

\section{Discussion}

Several approaches have been taken to estimating the predicted $\tau$ in M31
due to its own mass distribution.
Initially Crotts (1992) just approximated the entire mass of M31 as an
$r^{_{-2}}$ density distribution, which produces an optical depth for far-side
disk stars of $\tau \approx 10^{-5}$.
The presence of a core saturation radius will reduce $\tau$ in the center of
M31 while maintaining this high plateau value at larger radii.
Jetzer (1994) considers the effects of only the dark matter halo, with a large
core radius of 5~kpc, and finds a value of $\tau=1\times10^{-6}$ in the center
of M31, rising to $3\times10^{-6}$ at the outer edge of our field.
Han and Gould (1996) treat both the halo and bulge of M31 and find $\tau=
7\times10^{-6}$ in the center, dominated by the bulge, falling to
$3\times10^{-6}$ at the outside edge of our field, where the halo dominates.
There is a factor of about 1.3 disagreement between the lower values of Jetzer
versus Crotts and Han \& Gould due to different assumed values of M31's
rotation velocity.
Lensing of disk stars by other disk stars produces a $\tau$ component of
$4\times10^{-7}$ (Gould 1994), while a standard Galactic halo model adds
$\tau\approx1\times10^{-6}$ (Paczynski 1986).
Together, these components sum to at least $\tau\approx5\times10^{-6}$
throughout the field, which is the value that we will adopt for the sake of
discussion.
Note that this is about an order of magnitude greater than that suggested by
Galactic survey results towards the LMC ($\tau_{obs}=2.9^{+1.4}_{-0.9} \times
10^{-7}$, Alcock et al.~1996), or for predicted Galactic/LMC halo values
($\tau_{model}=4.7\times 10^{-7}$, Alcock et al.~1996), but only slightly
larger than Galactic Bulge results ($\tau_{obs}=(3.3 \pm 1.2) \times 10^{-6}$,
Paczynski et al.~1994; $\tau_{obs}=(3.9\pm 1.8)\times10^{-6}$, Alcock et
al.~1995).

From our previous constraints on the luminosity function of stars in our
field (TC), we have estimate that we are sensitive to detectable microlensing
of any of $6.9\times10^5$ stars in our field.
These data are primarily sensitive to timescales ranging from 2$^d$ to 10$^d$,
corresponding to 0.003~$M_\odot$ to 0.08~$M_\odot$.
We have 13 and 2 sample times respectively corresponding to $9.0\times 10^6$
and $1.4\times 10^6$ star-epochs.
The predicted number of events for this mass range given a $\tau_{Gal+M31}$
of $5\times 10^{-6}$ is 45 to 7 events.
Except for one possible detection at the upper end of this range, we find
no events on these timescales, thereby eliminating this mass range as a 100\%
contribution to the mass of M31 at considerably better than 95\% confidence.
On the other hand, we expect to detect approximately 2 events (given 100\%
efficiency) if the mass of M31 is made entirely of $1~M_\odot$ objects, while
we see six candidates, half of which are at this scale or larger.
This argues that some of these may not be caused by microlensing.

Our primary result is (1) the lack of any detection corresponding to masses up
to $0.08~M_\odot$ (with perhaps one exception), and (2) the possible detection
of events on the scale of about $1~M_\odot$.
The number of such events on this larger timescale, however, is significantly
greater than would be predicted given models of the lensing optical depth, so
might indicate contamination by variable stars.
Both of these results are consistent with microlensing searches in both the
Bulge and halo of our Galaxy (e.g. Paczynski et al.~1994, Alcock et al.~1995,
1996) in which few, if any, substellar masses are detected.
Likewise, slightly sub-solar masses are indicated as the primary cause of
microlensing events both towards the LMC and Bulge.
A further season of data will determine whether our six candidates are simply
variable stars, or exhibit constant baseline, implicating them as microlensing
events.

\acknowledgments

We appreciate the assistance of Robert Uglesich in the preparation of this
paper.
We also thank the Vatican Observatory Research Group, in particular Richard
Boyle and Chris Corbally, in assisting our efforts on the VATT, and the Steward
Observatory engineering group, in particular Richard Cromwell and Steve West.
Ata Sarajedini aided our program at the KPNO 4m.
This research was funded by the David and Lucile Packard Foundation.

\clearpage

\newpage 

\noindent Figure 1:
The left panel shows reduced but unsubtracted subimage (about 450 arcsec$^2$)
prior to DIP.
(Actually, the average of our 24 VATT epochs from autumn 1995 is shown.)~
Some of the sources shown here are variable, but at a level far below the
average surface brightness fluctuation in the image and below a level which can
be distinguished by eye.
The right panel shows the difference image of the same region, after a
high-$S/N$ sum of images has been scaled, registered, PSF adjusted and
subtracted using DIP, for UT 17 Oct 1995.
Several variable sources are revealed by the image subtraction, some fainter
than the average (black), some brighter (white).
The circled source corresponds to the fourth microlensing candidate shown
in Figure 2.

\noindent Figure 2:
the lightcurves of the six candidate microlensing events described in the text
and Table 1.
One count is equivalent to $R = 31.09$.

\noindent Figure 3:
the example described in the text of one of several lightcurves well-fit in the
peak by a microlensing model, but which does not maintain a consistent baseline
outside of the peak.

\newpage 

\begin{deluxetable}{cccccccc}
\tablecaption{Candidate M31 Microlensing Events}
\tablewidth{0pt}
\tablehead{
\colhead{RA} &
\colhead{Dec} &
\colhead{Minor Axis} &
\colhead{Duration,} &
\colhead{Probable} &
\colhead{Impact} &
\colhead{Baseline} &
\colhead{$\chi^2/\nu$} \\
\colhead{(J2000)} &
\colhead{(J2000)} &
\colhead{Distance,} &
\colhead{$t_e$ (days)} &
\colhead{Mass,} &
\colhead{Parameter} &
\colhead{Magn.,} &
\colhead{} \\
\colhead{} &
\colhead{} &
\colhead{$d$ (kpc)} &
\colhead{} &
\colhead{$m$ ($M_\odot$)} &
\colhead{vs.~$R_e$, $u_o$} &
\colhead{$R$} &
\colhead{}
} 
\startdata
0$^h42^m55^s$.7 & $+41^\circ14^\prime27^{\prime\prime}$ & 0.59 & ~7.8 & (0.09)$^a$ & 0.648 & 21.08 & 0.75 \nl 
0$^h42^m42^s$.3 & $+41^\circ11^\prime~2^{\prime\prime}$ & 0.62 & 55.7 & (4.3)$^a$  & 0.369 & 21.14 & 1.67 \nl 
0$^h42^m54^s$.1 & $+41^\circ10^\prime55^{\prime\prime}$ & 1.02 & 32.6 &  0.90$^b$  & 0.680 & 20.22 & 1.35 \nl 
0$^h43^m14^s$.8 & $+41^\circ12^\prime32^{\prime\prime}$ & 1.49 & 39.3 &  0.90$^b$  & 0.501 & 20.77 & 1.62 \nl 
0$^h43^m22^s$.6 & $+41^\circ~5^\prime52^{\prime\prime}$ & 2.67 & 31.7 &  0.32      & 0.590 & 20.93 & 2.23 \nl 
0$^h43^m49^s$.0 & $+41^\circ11^\prime28^{\prime\prime}$ & 2.77 & 39.9 &  0.49      & 0.690 & 20.61 & 2.01 \nl 
\enddata
\tablenotetext{a}{Source probably in bulge, so mass estimate is unreliable.}
\tablenotetext{b}{Source in bulge or disk; mass estimate assumes disk.}
\end{deluxetable}
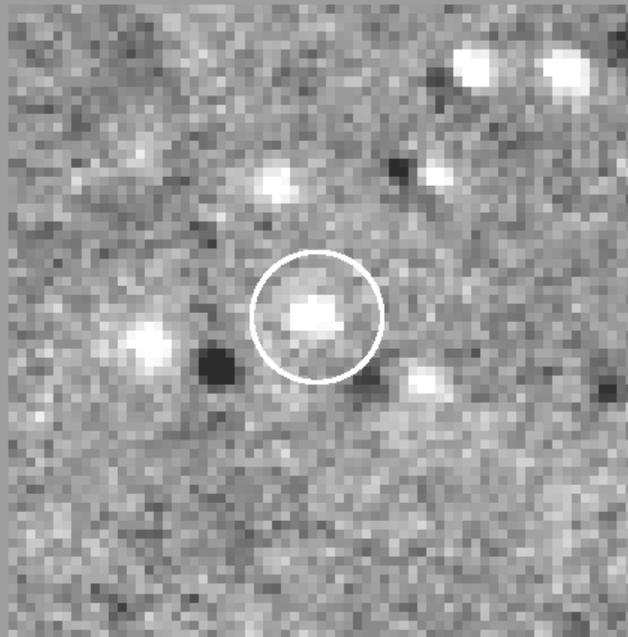
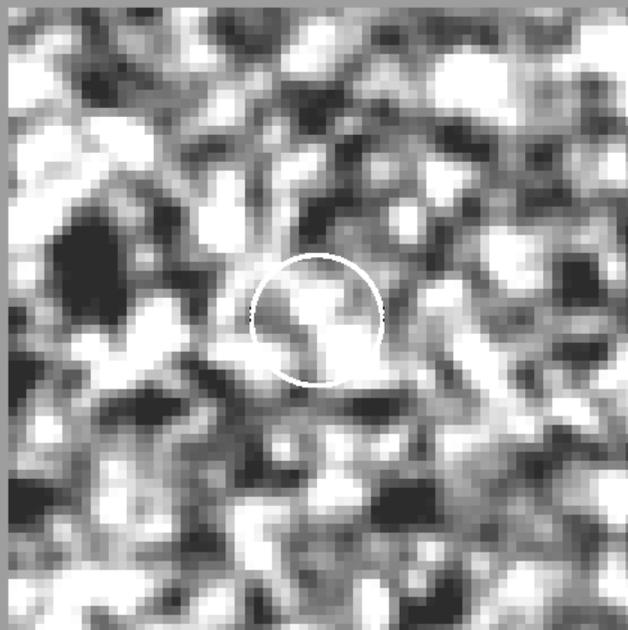
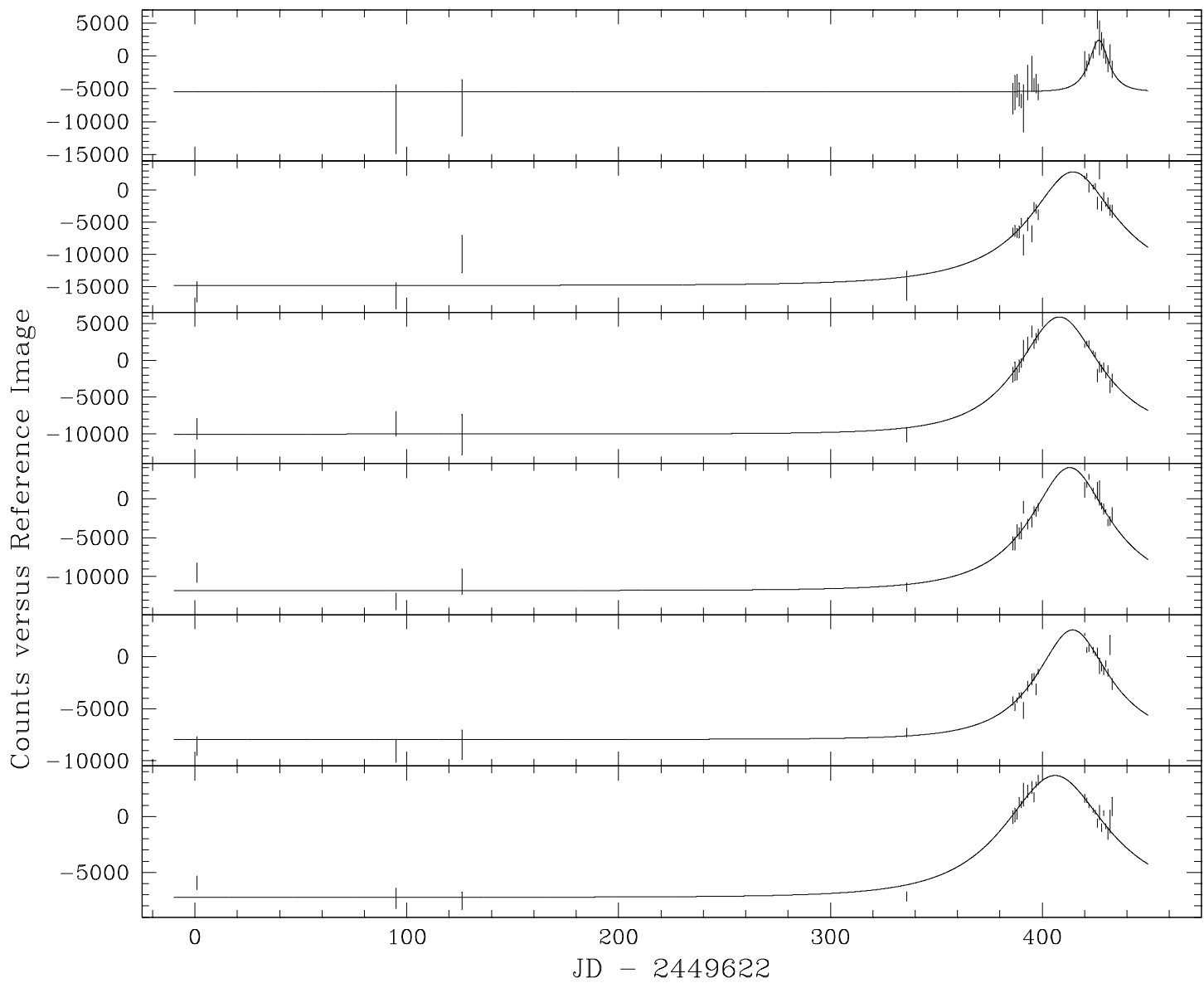
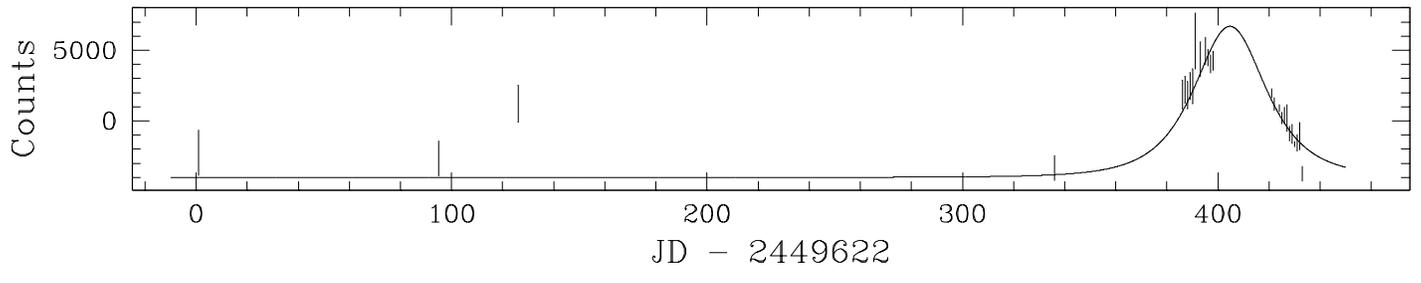
\end{document}